\documentclass[aps,pra,preprint,showpacs,showkeys,amsmath,amssymb]{revtex4-1}
\usepackage{graphicx}
\usepackage{dcolumn}
\usepackage{bm}
\usepackage{hyperref}
\begin{document}

\title{Local quantum uncertainty for multipartite quantum systems}
\author{Mazhar Ali}
\affiliation{Department of Electrical Engineering, Faculty of Engineering, Islamic University Madinah, 107 Madinah, Saudi Arabia}

\begin{abstract}
Local quantum uncertainty captures purely quantum correlations excluding their classical counterpart. This measure is quantum discord type, however 
with the advantage that there is no need to carry out the complicated optimization procedure over measurements. This measure is initially defined 
for bipartite quantum systems and a closed formula exists only for $2 \otimes d$ systems. We extend the idea of local quantum uncertainty to 
multi-qubit systems and provide the similar closed formula to compute this measure. We explicitly calculate local quantum uncertainty for various 
quantum states of three and four qubits, like GHZ state, W state, Dicke state, Cluster state, Singlet state, and Chi state all mixed with white noise. 
We compute this measure for some other well known three qubit quantum states as well. We show that for all such symmetric states, it is sufficient 
to apply measurements on any single qubit to compute this measure, whereas in general one has to apply measurements on all parties as local quantum 
uncertainties for each bipartition can be different for an arbitrary quantum state.
\end{abstract}



\maketitle

Quantum states are fundamentally different then classical states in such a way that any local measurements on one part of either bipartite or 
multipartite states necessarily give rise to uncertainty in results. This randomness is not a fault of measuring device but an integral nature of 
quantum states. Quantum entanglement, quantum nonlocality, and quantum discord are few quantitative manifestation of this randomness. 
The only states which are invariant under such local measurements are those states which can be described by classical probability distribution. 
Such states have zero quantum discord \cite{Ollivier-PRL88-2001, Henderson-JPA34-2001, Modi-RMP84-2012}. 
Quantum states for two or more parties may be entangled, however entanglement is not the only quantum correlation present among quantum states. 
There are quantum states which are separable, nevertheless quantum correlated (nonzero quantum discord). 
Quantum discord may be defined as the difference between quantum mutual information and classical 
correlations \cite{Ollivier-PRL88-2001, Henderson-JPA34-2001, Modi-RMP84-2012,Luo-PRA77-2008, Ali-PRA81-2010}. 
Due to complicated minimization process, the computation of quantum discord is not an easy task and analytical results are known only for some 
restricted families of states \cite{Ali-JPA43-2010, Rau-2017}. 
For $2 \otimes d$ quantum systems, analytical results for quantum discord are known for a specific family of states \cite{Ali-JPA43-2010} and 
the general procedure to calculate discord is also worked out \cite{Rau-2017}.
Some authors have proposed quantum discord for multipartite systems \cite{Mahdian-EPJD-2012, Xu-PLA-2013, Bai-PRA88-2013, Daoud-2012, Buscemi-PRA-2013}. 
Some other measures of such non-classical correlations include quantum work deficit \cite{Oppenheim-PRL89-2002}, quantum 
deficit \cite{Rajagopal-PRA66-2002}, measurement-induced non-locality \cite{Luo-PRL106-2011}, etc (see references in \cite{Bera-RPP81-2018}). 
The quantum correlations have utilization in potential applications, including remote state preparation \cite{Dakic-Nat-2012}, 
entanglement distribution \cite{Streltsov-PRL108-2012, Chuan-PRL109-2012}, transmission of correlations \cite{Streltsov-PRL111-2013}, and 
quantum meteorology \cite{Modi-PRX1-2011} to name few. It is in general hard task to characterize and quantify quantum correlations.  
Several authors have proposed different techniques to compute quantum correlations. The theory of quantum correlations have attracted lot of 
interest and considerable efforts have been devoted to it \cite{Horodecki-RMP-2009, gtreview, Chiara-RPP2018}. 

Recently, a discord-like measure has been proposed, known as local quantum uncertainty \cite{Girolami-PRL110-2013}. 
This measure is quantified via skew information which is achievable on a single local measurement \cite{Luo-PRL91-2003}.   
This measure has a closed formula calculated for $2 \otimes d$ bipartite quantum systems. Later on, some authors tried to study local quantum 
uncertainty for orthogonally invariant class of states \cite{Sen-QIP-2015}. This measure was also study for quantum phase 
transitions \cite{Coulamy-PLA-2016}. The relationship between local quantum uncertainty and quantum Fisher information under non-Markovian environment 
was also discussed \cite{Wu-AP-2018}. Recently, some authors have studied local quantum uncertainty under various decoherence models and also 
worked out some preliminary results for three qubits \cite{Slaoui-2019}. 
We extend local quantum uncertainty for multi-qubit quantum system. 
As there are several bipartition for multi-qubit system, we can define local quantum uncertainty for each bipartition. After  
calculating all such local quantum uncertainties, we suggest an arithmetic mean to calculate the average local quantum uncertainty for a given 
multi-qubit state. nevertheless, we find that for all specific quantum states which 
we study here, each local quantum uncertainty for every bipartition is exactly same due to symmetry of these quantum states. 
However, by taking a random state, we explicitly demonstrate that local quantum uncertainty can have a different value for each bipartition, so the 
average value gives local quantum uncertainty for given quantum state. We calculate this measure for various well known families of quantum states 
for three and four qubits and obtain analytical results. The benefit of this measure and its extension to multi-qubit has the advantage that we do not 
needs any complicated maximization or minimization over parameters related with measurements as one has to do to calculate quantum discord. 
Interestingly, for four qubits, we find that except $W$-states mixed with white noise, all other specific quantum states have same expressions for local 
quantum uncertainty.   

Local quantum uncertainty is a measure of quantum correlations which captures purely quantum part in a given quantum state by applying 
local measurements on one part of quantum state. This measure has been defined recently for $2 \otimes d$ quantum systems \cite{Girolami-PRL110-2013}. 
It is a quantum discord-type measure and for certain quantum states, quantum discord and local quantum uncertainty captures precisely same 
correlations and are equal to each other, whereas for some other states, they are different measures. The advantage of local quantum 
uncertainty over quantum discord is the fact that to compute local quantum uncertainty we only need to find the maximum eigenvalue of a 
symmetric $3 \times 3$ matrix. This is quite easy task as compared with complicated minimization procedure over parameters related with measurements. 
Local quantum uncertainty is defined as the minimum skew information which is obtained via local measurement on 
qubit part only, that is,  
\begin{equation}
\mathcal{Q}(\rho) \equiv \, \min_{K_A} \, \mathcal{I} (\rho , K_A \otimes \mathbb{I}_B) \,, 
\end{equation}
where $K_A$ is a hermitian operator (local observable) on subsystem $A$, and $\mathcal{I}$ is the skew information \cite{Luo-PRL91-2003} of the density 
operator $\rho$, defined as 
\begin{equation}
\mathcal{I} (\rho , K_A \otimes \mathbb{I}_B) \, = \,- \frac{1}{2} \, \rm{Tr} ( \, [ \, \sqrt{\rho}, \, K_A \otimes \mathbb{I}_B ]^2 \, ) \,.
\end{equation}
The skew information is nonnegative, and non-increasing under classical mixing. It has been shown \cite{Girolami-PRL110-2013} that 
for $2 \otimes d$ quantum systems, the compact formula for local quantum uncertainty is given as
\begin{equation}
\mathcal{Q}(\rho) = 1 - \rm{max} \, \{ \lambda_1 \,, \lambda_2 \, , \lambda_3 \, \}\,, 
\end{equation}
where $\lambda_i$ are the eigenvalues of $3 \times 3$ symmetric matrix $\mathcal{M}$. The matrix elements of symmetric matrix $\mathcal{M}$ 
are calculated by the relationship
\begin{equation}
 m_{ij} \equiv \rm{Tr} \, \big\{ \, \sqrt{\rho} \, (\sigma_i \otimes \mathbb{I}_B) \, \sqrt{\rho} \, (\sigma_j \otimes \mathbb{I}_B) \, \big\}\,, 
\end{equation}
where $i,j = 1,2,3$ and $\sigma_i$ are the standard Pauli matrices.

We generalize this definition of local quantum uncertainty for multi-qubit quantum systems as follows. First, we observe that 
the definition of local quantum uncertainty for $2 \otimes d$ systems can be applied to multi-qubit systems without any technical 
consequences because we can always regard multi-qubit system as $2 \otimes d$ systems, where $d = 2 \otimes 2 \otimes \ldots \otimes 2$ 
may represent the remaining $N-1$ qubits as $d$ dimensional quantum system. However, we note that multi-qubit systems have richer structure 
as compared with bipartite quantum systems. It might be the case that some bipartition are quantum correlated and some may be classically 
correlated. So we need to apply the local measurements across each bipartition in order to capture quantum correlations. 
To this aim, let $\rho$ be an arbitrary density matrix for $N$ qubits. We can apply the local measurements on each qubit $A$, $B$, $\ldots$, 
$N$. When we apply measurements on qubit $A$, we regard all rest of the qubits as $d$-dimensional system. Thus we obtain $N$ symmetric matrices.
For each bipartition, the matrix elements belonging to these $N$ symmetric matrices are calculated according to relations 
\begin{eqnarray}
\tilde{m}_{ij}^A =  \rm{Tr} \, \{ \, \sqrt{\rho} \, (\sigma_i \otimes \mathbb{I}_2 \otimes \ldots \otimes \mathbb{I}_2) \, \nonumber \\ 
\times \, \sqrt{\rho} \, (\sigma_j \otimes \mathbb{I}_2 \otimes \ldots \otimes \mathbb{I}_2) \, \}\,, \nonumber \\ 
\tilde{m}_{ij}^B = \rm{Tr} \, \{ \, \sqrt{\rho} \, (\mathbb{I}_2 \otimes \sigma_i \otimes \ldots \otimes \mathbb{I}_2) \, \nonumber \\ 
\times \, \sqrt{\rho} \, (\mathbb{I}_2 \otimes \sigma_j \otimes \ldots \otimes \mathbb{I}_2) \, \}\,, \nonumber \\
\vdots \nonumber \\
\vdots \nonumber \\
\tilde{m}_{ij}^N = \rm{Tr} \, \{ \, \sqrt{\rho} \, (\mathbb{I}_2 \otimes \mathbb{I}_2 \ldots \otimes \sigma_i) \, \nonumber \\ 
\times \, \sqrt{\rho} \, (\mathbb{I}_2 \otimes \mathbb{I}_2 \otimes \ldots \otimes \sigma_j) \}\,. 
\end{eqnarray}
The corresponding eigenvalues of such $3 \times 3$ symmetric matrices $\mathcal{\tilde{M}}_i$ can be determined easily. 
The local quantum uncertainties related with each bipartition are defined as follows
\begin{eqnarray}
\mathcal{Q}_{A/BC\ldots N}(\rho) = 1 - \rm{max} \, \{ \text{Spectrum of } \mathcal{\tilde{M}}_A \, \}\,\nonumber \\ 
\mathcal{Q}_{B/AC\ldots N}(\rho) = 1 - \rm{max} \, \{ \text{Spectrum of } \mathcal{\tilde{M}}_B \, \}\,\nonumber \\ 
\vdots \nonumber \\
\vdots \nonumber \\
\mathcal{Q}_{N/ABC\ldots N-1}(\rho) = 1 - \rm{max} \, \{ \text{Spectrum of } \mathcal{\tilde{M}}_N \, \}\,.
\end{eqnarray}
Finally we propose the mean value of local quantum uncertainty for a given $N$-qubits quantum state to be calculated as
\begin{equation}
\mathcal{Q}(\rho_{N}) = \frac{\sum_{i = A}^N \, \mathcal{Q}_{i/N_i}}{N}\,, 
\end{equation}
where $N_i$ are the remaining $N-1$ qubits except $i$. 

As a concrete example, let us consider the case of three qubits. Following the procedure mentioned above, we can find $\mathcal{\tilde{M}}_A$ 
for bipartition $A/BC$, $\mathcal{\tilde{M}}_B$ for bipartition $B/CA$, 
and $\mathcal{\tilde{M}}_C$ for bipartition $C/AB$. The respective matrix elements are calculated using relations
\begin{equation}
\tilde{m}_{ij}^A = \rm{Tr} \, \{ \, \sqrt{\rho_{ABC}} \, (\sigma_i \otimes \mathbb{I}_2 \otimes \mathbb{I}_2) \, 
\sqrt{\rho_{ABC}} \, (\sigma_j \otimes \mathbb{I}_2 \otimes \mathbb{I}_2) \, \}\,, 
\end{equation}
\begin{equation}
\tilde{m}_{ij}^B = \rm{Tr} \, \{ \, \sqrt{\rho_{ABC}} \, (\mathbb{I}_2 \otimes \sigma_i \otimes \mathbb{I}_2) \, 
\sqrt{\rho_{ABC}} \, (\mathbb{I}_2 \otimes \sigma_j \otimes  \mathbb{I}_2) \, \}\,, 
\end{equation}
\begin{equation}
\tilde{m}_{ij}^C = \rm{Tr} \, \{ \, \sqrt{\rho_{ABC}} \, (\mathbb{I}_2 \otimes \mathbb{I}_2 \otimes \sigma_i ) \, 
\sqrt{\rho_{ABC}} \, (\mathbb{I}_2 \otimes \mathbb{I}_2 \otimes \sigma_j) \}\,, 
\end{equation}
where $\tilde{m}_{ij}^A \neq \tilde{m}_{ij}^B \neq \tilde{m}_{ij}^C$ in general, however they may be equal to each other for some special cases. 
We mention here that the number of these symmetric matrices are same as the number of qubits. 
The local quantum uncertainty in this situation would be defined as 
\begin{equation}
\mathcal{Q}(\rho_{ABC}) = \frac{\big( \mathcal{Q}_{A/BC} \, + \, \mathcal{Q}_{B/CA} \, + \, \mathcal{Q}_{C/AB} \big)}{3}\,, 
\end{equation}
where
\begin{equation}
\mathcal{Q}_{A/BC}  = 1 - \max \, \big\{ \, \{ \mathcal{\tilde{M}}_A \}\, \big\}, 
\end{equation}
\begin{equation}
\mathcal{Q}_{B/CA}  = 1 - \max  \, \big\{ \, \{ \mathcal{\tilde{M}}_B \}\, \big\} \,, 
\end{equation}
\begin{equation}
\mathcal{Q}_{C/AB}  = 1 - \max  \, \big\{ \, \{ \mathcal{\tilde{M}}_C \}\, \big\} \,, 
\end{equation}
where $\{ \mathcal{\tilde{M}}_i \}$ denote the spectrum (eigenvalues) of the corresponding $3 \times 3$ matrix $\mathcal{\tilde{M}}_i$. 
For the special case when all three matrices have the same set of eigenvalues then $\mathcal{Q}_{A/BC} = \mathcal{Q}_{B/CA} = \mathcal{Q}_{C/AB}$ 
and $\mathcal{Q}(\rho_{ABC}) = \mathcal{Q}_{A/BC}$. In this case measurements need to be applied to any one qubit. 

We will now present some examples computing local quantum uncertainty for various families of three qubits and 
four qubits quantum states. An important family of quantum states is GHZ states mixed with white noise. These states for 
three qubits are defined as
\begin{eqnarray}
\rho_{GHZ_3} = (1-\alpha) \, | GHZ_3 \rangle\langle GHZ_3 | + \frac{\alpha}{8} \, \mathbb{I}_8 \,,
\label{Eq:ghz3}
\end{eqnarray}
where $ 0 \leq \alpha \leq 1$, $\mathbb{I}_8/8$ is maximally mixed state, and maximally entangled pure state is given as 
\begin{eqnarray}
| GHZ_3 \rangle = \frac{1}{\sqrt{2}} \, ( \, |0 0 0 \rangle + |1 1 1 \rangle \, ) \, .  
\end{eqnarray}
Entanglement properties of these states Eq.~(\ref{Eq:ghz3}) are well known \cite{Guehne-2010}. It is known that these states are fully 
separable for $ 0.8 \leq \alpha \leq 1$, bi-separable for $0.571 \leq \alpha < 0.8$, and genuine entangled for 
$0 \leq \alpha < 0.571$ \cite{Guehne-2010}. 
We have calculated all three symmetric matrices $\mathcal{\tilde{M}}_i$ for measurements on qubit $A$, $B$, and $C$. It turns out that all 
three matrices are same and therefore have the same set of three eigenvalues. In addition, all three eigenvalues are also same, so the 
problem to pick the maximum eigenvalue is even trivial. The maximum eigenvalue is given as 
\begin{eqnarray}
\lambda = \frac{3 \, \alpha  + \sqrt{\alpha (8 - 7 \, \alpha)}}{4} \,.
\end{eqnarray}
Therefore, local quantum uncertainty for states Eq.~(\ref{Eq:ghz3}) is simply 
\begin{eqnarray}
\mathcal{Q} (\rho_{GHZ_3}) = 1 - \frac{3 \, \alpha  + \sqrt{\alpha (8 - 7 \, \alpha)}}{4} \,.
\end{eqnarray}
We observe that for $\alpha = 0$, $ \mathcal{Q} (\rho_{GHZ_3}) = 1$ which is expected as pure maximally entangled state has maximum correlations. 
We note that for $\alpha = 1$, we have $ \mathcal{Q} (\rho_{GHZ_3}) = 0$, which is also expected result because maximally mixed state is classically 
correlated and have no quantum correlations in it. For other values of $\alpha < 1$, local quantum uncertainty $\mathcal{Q}(\rho_{GHZ_3}) > 0$. 
We have seen that local quantum uncertainty precisely captures quantum correlations just like quantum discord.  

Second important class of states for three qubits is $W$ state mixed with while noise. These states are defined as
\begin{eqnarray}
\rho_{W_3} = (1-\beta) \, | \, W_3 \, \rangle\langle \, W_3 \, | + \frac{\beta}{8} \, \mathbb{I}_8 \,,
\label{Eq:W3}
\end{eqnarray}
where $ 0 \leq \beta \leq 1$ and $W_3$ state is given as
\begin{eqnarray}
| \, W_3 \, \rangle = \frac{1}{\sqrt{3}} \, ( \, |0 0 1 \rangle + |0 1 0\rangle + |1 0 0\rangle\, ) \, .  
\end{eqnarray}
The entanglement properties of Eq.~(\ref{Eq:W3}) are also well known. These states are fully separable or bi-separable for $0.521 \leq \beta \leq 1$, 
whereas genuine tripartite entangled for $ 0 \leq \beta < 0.521$ \cite{Guehne-2010}. 
We now carry out the same procedure as mentioned earlier to compute local quantum uncertainty. We calculated all three symmetric 
matrices and found them to be exactly equal to each other as it was the case for $\rho_{GHZ_3}$ states. Therefore, we get same 
set of three eigenvalues for all three bipartition. Two of the eigenvalues are equal to each other, whereas third eigenvalue is 
different. These eigenvalues are given as 
\begin{eqnarray}
w_1 = w_2 = \frac{3 \, \beta + \sqrt{\beta (8 - 7 \, \beta)}}{4} \, ,\nonumber \\
w_3 = \frac{1 + 6 \, \beta + 2 \, \sqrt{\beta (8 - 7 \, \beta) }}{9}\, .  
\end{eqnarray}
It is not difficult to check that $w_3 > w_1$, for all values of parameter $\beta$. The local quantum uncertainty for states Eq.~(\ref{Eq:W3}) 
is simply given as
\begin{eqnarray}
\mathcal{Q} (\rho_{W_3}) = \frac{8 - 6 \, \beta - 2 \, \sqrt{\beta (8 - 7 \, \beta)}}{9} \,.
\end{eqnarray}
We can readily check that for $\beta = 0$, we get $ \mathcal{Q} (\rho_{W_3}) = 8/9$. This means that for pure $W_3$ state, local quantum uncertainty 
does not have maximum value of $1$. The genuine negativity for $W_3$ state is also not maximum, whereas $GHZ$ state is regarded as maximally 
entangled as measured by genuine negativity \cite{Guehne-2010}. We can also check that for $\beta = 1$, we have $ \mathcal{Q} (\rho_{W_3}) = 0$ as it 
should be. 

Let us take another example of three qubits quantum states defined as
\begin{widetext}
\begin{equation}
\rho_{AK} = \frac{1}{8 + 8 \, \gamma} \, \left(
\begin{array}{llllllll}
4 +\gamma & 0 & 0 & 0 & 0 & 0 & 0 & 2 \\
0 & \gamma & 0 & 0 & 0 & 0 & 2 & 0 \\
0 & 0 & \gamma & 0 & 0 & -2 & 0 & 0 \\
0 & 0 & 0 & \gamma & 2 & 0 & 0 & 0 \\
0 & 0 & 0 & 2 & \gamma & 0 & 0 & 0 \\
0 & 0 & -2 & 0 & 0 & \gamma & 0 & 0 \\
0 & 2 & 0 & 0 & 0 & 0 & \gamma & 0 \\
2 & 0 & 0 & 0 & 0 & 0 & 0 & 4 + \gamma
\end{array}
\right) \,.\label{Eq:INF}
\end{equation}
\end{widetext}
This matrix is a valid quantum state for $\gamma \geq 2$. This family of states may be called Kay states as they were introduced 
by A. Kay \cite{Kay-PRA83-2011}. The states have positive partial transpose (PPT) with respect of all bipartition. It is known that 
for $ 2 \leq \gamma < 2 \, \sqrt{2}$, this density matrix is bound entangled and for $\alpha \geq 2 \, \sqrt{2}$, 
the state is separable. 
We calculate all three symmetric matrices $\mathcal{\tilde{M}}_i$ for this state and find that once again, they are equal to each other. 
There are two eigenvalues which are same whereas the third eigenvalue is different. These eigenvalues are given as
\begin{eqnarray}
k_1 = k_2 = \frac{1}{4} \, \sqrt{\frac{\gamma + 2}{\gamma + 1}} \, \bigg( 3 \sqrt{\frac{- 2 + \gamma}{1 + \gamma}} 
+ \sqrt{\frac{6 + \gamma}{1 + \gamma}} \, \bigg) \, ,\nonumber \\
k_3 = \frac{3 \, \gamma + 2 + \sqrt{(\gamma - 2)(6 + \gamma)}}{4 (\gamma +1)}\, .  
\end{eqnarray}
It is not difficult to find that $k_3 > k_1$, therefore local quantum uncertainty for Kay-states is given as
\begin{eqnarray}
\mathcal{Q} (\rho_{AK}) = \frac{2 + \gamma - \sqrt{(\gamma -2 )(6 + \gamma)}}{4 (1 + \gamma)} \,.
\end{eqnarray}
This expression is not real for $\gamma < 2$, so local quantum uncertainty also reflects this restriction on parameter in quantum states.

\begin{figure}[t!]
\includegraphics{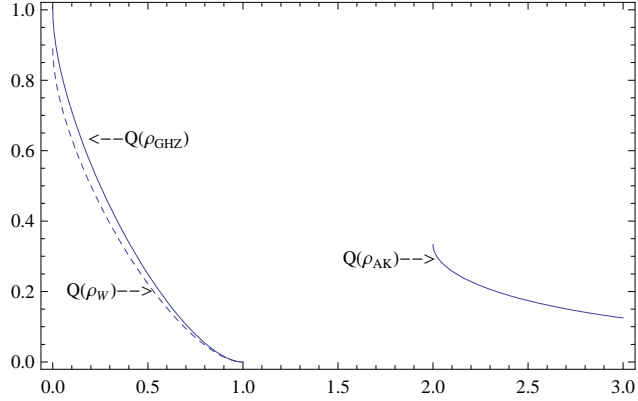}
\caption{Local quantum uncertainty is plotted again single parameter for $\rho_{GHZ}$, $\rho_{W_3}$, and $\rho_{AK}$ states.  
See text for explanations.}
\label{Fig:1}
\end{figure}
Figure~(\ref{Fig:1}) shows local quantum uncertainty $\mathcal{Q} (\rho)$ plotted against corresponding single parameter for Eq.~(\ref{Eq:ghz3}), 
Eq.~(\ref{Eq:W3}), and Eq.~(\ref{Eq:INF}). The quantum correlations in GHZ and W state are highest for pure states and as mixing increases, 
the correlations decrease and finally become zero for maximally mixed state. For Kay-states the quantum correlations are largest for $\gamma = 2$, 
which is a bound entangled state. As we increase the value of parameter, quantum states move towards separable states and quantum correlations are 
smaller than the bound entangled state. We have checked local quantum uncertainty even for very large values of parameter $\gamma $ and found 
local quantum uncertainty still strictly greater than zero.

In all above examples, we have seen that local quantum uncertainty for each bipartition turns out to be same. However, it is not true for 
the set of all quantum states as there exist other states for which each bipartition may have different local quantum uncertainty. 
We demonstrate this difference simply by taking a random state and calculating local quantum uncertainty for each bipartition. To this aim, 
first we generate a random pure state and then mix it with white noise such that white noise fraction is $0.2$ and random state fraction is $0.8$. 
The corresponding symmetric matrix with measurements on 
qubit $A$ is given as
\begin{eqnarray}
\mathcal{\tilde{M}}_A \approx \left(
\begin{array}{lll}
0.65 & 0.014 & 0.115 \\
0.014 & 0.594 & -0.015 \\
0.115 & -0.015 & 0.757
\end{array}
\right) \,,\label{Eq:rnd1}
\end{eqnarray}
with the eigenvalues $(\, 0.83, \, 0.61, \, 0.56 \, )$. For measurements on qubit $B$, we get   
\begin{eqnarray}
\mathcal{\tilde{M}}_B \approx \left(
\begin{array}{lll}
0.59 & 0.01 & -0.05 \\
0.01 & 0.651 & 0.107 \\
-0.05 & 0.107 & 0.687
\end{array}
\right) \,, \label{Eq:rnd2}
\end{eqnarray}
with eigenvalues $ (\, 0.78, \, 0.61, \, 0.53 \, )$, and finally for qubit $C$, we have 
\begin{eqnarray}
\mathcal{\tilde{M}}_C \approx \left(
\begin{array}{lll}
0.63 & -0.112 & -0.032 \\
-0.112 & 0.83 & 0.12 \\
-0.032 & 0.12 & 0.711
\end{array}
\right) \,,\label{Eq:rnd3}
\end{eqnarray}
with eigenvalues $ (\, 0.94, \, 0.65, \, 0.57 \, )$. The respective local quantum uncertainties are $\mathcal{Q}(\rho_{A/BC}) \approx 0.17$, 
$\mathcal{Q}(\rho_{B/CA}) \approx 0.22$, and $\mathcal{Q}(\rho_{C/AB}) \approx 0.06$. The average value is $\mathcal{Q}(\rho) \approx 0.15$.

Let us consider few examples of four qubit quantum states. Two important quantum states for four qubits are the GHZ state and W state given as  
\begin{eqnarray}
|GHZ_4 \rangle = \frac{1}{\sqrt{2}}(| 0 0 0 0 \rangle + |1 1 1 1\rangle ), \nonumber \\
|W_4 \rangle = \frac{1}{2}(|0 0 0 1\rangle + |0 0 1 0\rangle + |0 1 0 0 \rangle + |1 0 0 0 \rangle).
\label{Eq:GHZW}
\end{eqnarray}
For the GHZ state, the entanglement monotone has a value of $E(|GHZ_4\rangle\langle GHZ_4|) = 1$, while for the W state, its 
value is $E(|W_3\rangle\langle W_3|) \approx 0.886$ and $E(|W_4\rangle\langle W_4|) \approx 0.732$.

Several other four qubit quantum states are interesting and have been discussed in the literature. These states are the 
Dicke state $|D_{2,4}\rangle$, the four-qubit singlet state $|\Psi_{S,4}\rangle$, the cluster state $|CL\rangle$ and the so-called 
$\chi$-state $|\chi_4\rangle$. These quantum states are explicitly given as
\begin{eqnarray}
|D_{2,4} \rangle = \frac{1}{\sqrt{6}} \, \big[ |0011 \rangle + |1100\rangle + |0101 \rangle + |0110\rangle \nonumber \\ 
+ |1001 \rangle + | 1010\rangle \big] \,, \nonumber \\
|\Psi_{S,4}\rangle = \frac{1}{\sqrt{3}} \, \big[ |0011 \rangle + |1100\rangle - \frac{1}{2} \{ \, |0101 \rangle  
+ |0110\rangle  \nonumber \\ + |1001 \rangle + |1010\rangle \} \big] \,, \nonumber \\
|CL \rangle = \frac{1}{2} \, \big[|0000 \rangle + |0011\rangle + |1100 \rangle - |1111\rangle \big], \nonumber \\
|\chi_4 \rangle = \frac{1}{\sqrt{6}} \, \big\{\sqrt{2} \, | 1111 \rangle + |0001 \rangle + |0010\rangle 
+ |0100 \rangle \nonumber \\ + |1000\rangle \big\}. 
\end{eqnarray}
Note that all of these states have the maximum value of entanglement
$E (|D_{2,4} \rangle\langle D_{2,4}|) = E (|\Psi_{S,4} \rangle\langle \Psi_{S, 4}|) = E (|CL\rangle\langle CL|) 
= E (|\chi_4\rangle\langle \chi_4|) = 1$. Further entanglement properties of these states are reviewed in Ref.~\cite{gtreview}.

To find local quantum uncertainty, we first mix all of these states with white noise as follows
\begin{equation}
\rho_\eta =  (1 - \eta ) \, |\psi \rangle\langle \psi | + \frac{\eta}{16} \, \mathbb{I}_{16} \,,    
\label{Eq:mx}
\end{equation}
where $ 0 \leq \eta \leq 1$, and $|\psi\rangle$ is any of the above defined four qubit pure states. 
Next we calculate the four symmetric matrices for each one of these states and find out that they are all equal for every bipartition, 
that is, $\mathcal{\tilde{M}}_A = \mathcal{\tilde{M}}_B = \mathcal{\tilde{M}}_C = \mathcal{\tilde{M}}_D$. This implies that 
local quantum uncertainty for each bipartition is same.  Another 
interesting observation is that except $W_4$ state mixed with white noise, all other remaining five mixtures have exactly the 
same eigenvalues and consequently exactly the same expressions for local quantum uncertainty as well, that is, 
\begin{eqnarray}
\mathcal{Q}(\rho_{GHZ_4}) =  \mathcal{Q}(\rho_{D_{2,4}}) = \mathcal{Q}(\rho_{\Psi_{S, 4}}) \nonumber \\ 
=  \mathcal{Q}(\rho_{CL}) = \mathcal{Q}(\rho_{\chi_4}) = \mathcal{Q} (\rho_\eta) \,,     
\end{eqnarray}
where local quantum uncertainty for any of such states Eq.~(\ref{Eq:mx}) is given as 
\begin{equation}
\mathcal{Q}(\rho_\eta) = 1 - \frac{7 \, \eta + \sqrt{\eta (16 - 15 \, \eta)}}{8} \, . 
\end{equation}
We note that $\mathcal{Q}(\rho_\eta) = 1$ for $\eta = 0$, which means that GHZ state, Dicke State, singlet state, cluster state and chi 
state all have maximum amount of quantum correlations. We also note that $\mathcal{Q}(\rho_\eta) = 0$ for $\eta = 1$.  
For $W_4$ state mixed with white noise, local quantum uncertainty is given as
\begin{equation}
\mathcal{Q} (\rho_{W_4})= 1 - \frac{8 + 21 \, \eta + 3 \, \sqrt{\eta (16 - 15 \, \eta)}}{32} \, . 
\end{equation}
This value is $3/4 = 0.75$ for $\eta = 0$ and zero for $\eta = 1$. We have seen that for both $W_3$ and $W_4$ state, the numerical value of local 
quantum uncertainty is slightly larger than numerical value of genuine entanglement. 

We can easily demonstrate by generating a random state of four qubits that in general 
$\mathcal{Q}(\rho_{A/BCD}) \neq \mathcal{Q}(\rho_{B/CDA}) \neq \mathcal{Q}(\rho_{C/DBA}) \neq \mathcal{Q}(\rho_{D/ABC})$ as we have seen 
for three qubits.

In summary, we have extended the idea of local quantum uncertainty for multi-qubit quantum systems. We have analytically calculated this measure 
for several important families of quantum states of three and four qubits mixed with white noise. We find that all specific quantum states mixtures 
are symmetric as they all give the same value of local quantum uncertainty for measurements on each bipartition. Therefore for such states, measurements 
on any single qubit is sufficient to compute local quantum uncertainty. We have explicitly shown by taking a random state of three qubits 
that symmetric matrices resulting from measurements on each partition are not the same and hence the corresponding eigenvalues and local quantum 
uncertainties are also not equal to each other. Hence we get a different numerical value of local quantum uncertainty for each bipartition. 
Similar matrices should also be different for an arbitrary quantum state of four or higher number of qubits. This method is applicable to any 
arbitrary initial quantum state of $N$ qubits.  

\acknowledgements

The author is grateful to F. B. Abdullah for fruitful discussions.


\begin{thebibliography}{99}

\bibitem{Ollivier-PRL88-2001} H. Ollivier and W. H. Zurek, Phys. Rev. Lett. {\bf 88}, 017901 (2001).

\bibitem{Henderson-JPA34-2001} L. Henderson and V. Vedral, J. Phys. A {\bf 34}, 6899 (2001); V. Vedral, Phys. Rev. Lett. {\bf 90}, 050401 (2003).  

\bibitem{Modi-RMP84-2012} K. Modi, A. Brodutch, H. Cable, T. Paterek, and V. Vedral, Rev. Mod. Phys. {\bf 84}, 1655 (2012).

\bibitem{Luo-PRA77-2008} S. Luo, Phys. Rev. A {\bf 77}, 042303 (2008). 

\bibitem{Ali-PRA81-2010} M. Ali, A. R. P. Rau, and G. Alber, Phys. Rev A {\bf 81}, 042105 (2010); {ibid}, Phys. Rev. A {\bf 82}, 069902 (E) (2010); 
Y. Huang, Phys. Rev. A {\bf 88}, 014302 (2013).
%
\bibitem{Ali-JPA43-2010} M. Ali, J. Phys. A: Math. Theor. {\bf 43}, 495303 (2010).

\bibitem{Rau-2017} S. Vinjanampathy and A. R. P. Rau, J. Phys. A: Math. Theor. {\bf 45}, 095303 (2012); 
A. R. P. Rau, Quant. Info. Proc. {\bf 17} 216 (2018). 

\bibitem{Mahdian-EPJD-2012} M. Mahdian, R. Yousefjani, and S. Salimi, Eur. Phys. J. D {\bf 66}, 133 (2012).

\bibitem{Xu-PLA-2013} J. Xu, Phys. Lett. A {\bf 377}, 238 (2013).

\bibitem{Bai-PRA88-2013} Y-K. Bai, N. Zhang, M-Y. Ye, and Z. D. Wang, Phys. Rev. A {\bf 88}, 012123 (2013); H. C. Braga, C. C. Rulli, 
T. R. de Oliveira, and M. S. Sarandy, Phys. Rev. A {86}, 062106 (2012).

\bibitem{Daoud-2012} M. Daoud, and R. Ahl Laamara, Phys. Lett. A {\bf 376}, 2361 (2012); M. Daoud, and R. Ahl Laamara, 
Int. J. Qunt. Info. {\bf 10}, 1250060 (2012).

\bibitem{Buscemi-PRA-2013} F. Buscemi, and P. Bordone, Phys. Rev. A {\bf 87}, 042310 (2013); A. Beggi, F. Buscemi, and P. Bordone, 
Quan. Inf. Proc. {\bf 14}, 573 (2015); D. P. Chi, J. S. Kim, and K. Lee, 
Phys. Rev. A {\bf 87}, 062339 (2013); A. Y. Chernyavskiy, S. I. Doronin, and E. B. Feldman, Phys. Scr. T {\bf 160}, 014007 (2014).

\bibitem{Oppenheim-PRL89-2002} J. Oppenheim, M. Horodecki, P. Horodecki, and R. Horodecki, Phys. Rev. Lett. {\bf 89}, 180402 (2002); 
M. Horodecki, K. Horodecki, P. Horodecki, R. Horodecki, J. Oppenheim, A. Sen (De), and U. Sen, Phys. Rev. Lett. {\bf 90}, 100402 (2003);
M. Horodecki, P. Horodecki, R. Horodecki, J. Oppenheim, A. Sen (De), U. Sen, and B. Synak-Radtke, Phys. Rev. A {\bf 71}, 062307 (2005);
I. Devetak, Phys. Rev. A {\bf 71}, 062303 (2005).

\bibitem{Rajagopal-PRA66-2002} A. K. Rajagopal, and R. W. Rendell, Phys. Rev. A {\bf 66}, 022104 (2002); S. Luo, Phys. Rev. A {\bf 77}, 
022301 (2008).

\bibitem{Luo-PRL106-2011} S. Luo, and S. Fu, Phys. Rev. Lett. {\bf 106}, 120401 (2011).

\bibitem{Bera-RPP81-2018} A. Bera, T. Das, D.Sadhukhan, S. S. Roy, A. Sen De, and U. Sen, Rep. Prog. Phys. {\bf 81}, 024001 (2018). 


\bibitem{Dakic-Nat-2012} B. Dakic, {\it et al.}, Nat. Phys. {\bf 8}, 666 (2012).

\bibitem{Streltsov-PRL108-2012} A. Streltsov, H. Kempermann, and D. Bru\ss , Phys. Rev. Lett. {\bf 108}, 250501 (2012).

\bibitem{Chuan-PRL109-2012} T. K. Chuan, {\it et al.}, Phys. Rev. Lett. {\bf 109}, 070501 (2012).

\bibitem{Streltsov-PRL111-2013} A. Streltsov, and W. H. Zurek, Phys. Rev. Lett. {\bf 111}, 040401 (2013).

\bibitem{Modi-PRX1-2011} K. Modi, H. Cable, M. Williamson, and V. Vedral, Phys. Rev. X {\bf 1}, 021022 (2011).

\bibitem{Horodecki-RMP-2009} R. Horodecki {\it et al.}, {Rev. Mod. Phys.} {\bf 81}, 865 (2009).

\bibitem{gtreview} O. G\"uhne and G. T\'oth, {Phys. Rep.} {\bf 474}, 1 (2009).

\bibitem{Chiara-RPP2018} G. D. Chiara, and A. Sanpera, Rep. Prog. Phys. {\bf 81}, 074002 (2018). 

\bibitem{Girolami-PRL110-2013} D. Girolami, T. Tufarelli, and G. Adesso, Phys. Rev. Lett. {\bf 110}, 240402 (2013).

\bibitem{Luo-PRL91-2003} S. Luo, Phys. Rev. Lett. {\bf 91}, 180403 (2003); E. P. Wigner, and M. M. Yanase, 
Proc. Natl. Acad. Sci. {\bf 49}, 910 (1963); S. Luo, S. Fu, and C. H. Oh, Phys. Rev. A {\bf 85}, 032117 (2012).  

\bibitem{Sen-QIP-2015} A. Sen, A. Bhar, and D. Sarkar, Quan. Info. Proc. {\bf 14}, 269 (2015).

\bibitem{Coulamy-PLA-2016} I. B. Coulamy, J. H. Warnes, M. S. Sarandy, and A. Saguia, Phys. Lett. A {\bf 380}, 1724 (2016). 

\bibitem{Wu-AP-2018} S-X. Wu, Y. Zhang, and C-S. Yu, Ann. Phys. {\bf 390}, 71 (2018).

\bibitem{Slaoui-2019} A. Slaoui, M. I. Shaukat, M. Daoud, and R. Ahl Laamara, Eur. Phys. J. Plus {\bf 133}, 413 (2018);
A. Slaoui, M. Daoud, and R. Ahl Laamara, Quan. Info. Proc. {\bf 17}, 178 (2018); 
A. Slaoui, L. Bakmou, M. Daoud, and R. Ahl Laamara, Phys. Lett. A {\bf 383}, 2241 (2019); 
A. Slaoui, M. Daoud, and R. Ahl Laamara, Quan. Info. Proc. {\bf 18}, 250 (2019).

\bibitem{Guehne-2010} O. G\"uhne, and M. Seevinck, New. J. Phys. {\bf 12}, 053002 (2010); 
O. G\"uhne, Phys. Lett. A {\bf 375}, 406 (2011); 
B. Jungnitsch, T. Moroder, and O. G\"uhne, Phys. Rev. Lett. {\bf 106}, 190502 (2011);
L. Novo, T. Moroder, and O. G\"uhne, Phys. Rev. A {\bf 88}, 012305 (2013);
M. Hofmann, T. Moroder, and O. G\"uhne, J. Phys. A: Math. Theor. {\bf 47}, 155301 (2014).

\bibitem{Kay-PRA83-2011} A. Kay, Phys. Rev. A {\bf 83}, 020303 (R) (2011). 

\end{thebibliography}
\end{document}